\documentclass[12pt]{article}

\usepackage{graphicx}

\def\gtwid{\mathrel{\raise.3ex\hbox{$>$\kern-.75em\lower1ex\hbox{$\sim$}}}}
\def\ltwid{\mathrel{\raise.3ex\hbox{$<$\kern-.75em\lower1ex\hbox{$\sim$}}}}
\def\square{\kern1pt\vbox{\hrule height 1.2pt\hbox{\vrule width 1.2pt\hskip 3pt
   \vbox{\vskip 6pt}\hskip 3pt\vrule width 0.6pt}\hrule height 0.6pt}\kern1pt}

\begin{document}

\begin{titlepage}

\begin{flushright}
UFIFT-QG-16-04
\end{flushright}

\vskip 2cm

\begin{center}
{\bf Testing an Ansatz for the Leading Secular Loop Corrections from Quantum 
Gravity during Inflation}
\end{center}

\vskip 1cm

\begin{center}
S. Basu$^{*}$ and R. P. Woodard$^{\dagger}$
\end{center}

\vskip .5cm

\begin{center}
\it{$^{1}$ Department of Physics, University of Florida,\\
Gainesville, FL 32611, UNITED STATES}
\end{center}

\vspace{1cm}

\begin{center}
ABSTRACT
\end{center}
It is widely believed that the leading secular loop corrections from 
quantum gravity can be subsumed into a coordinate redefinition. Hence
the apparent infrared logarithm corrections to any quantity would be 
just the result of taking the expectation value of the tree order 
quantity at the transformed coordinates in the graviton vacuum. We
term this the {\it Transformation Ansatz} and we compare its predictions 
against explicit one loop computations in Maxwell + Einstein and Dirac + 
Einstein on de Sitter background. In each case the ansatz fails.

\begin{flushleft}
\end{flushleft}

\vskip .5cm

\begin{flushleft}
$^{*}$ e-mail: shinjinibasu@ufl.edu \\
$^{\dagger}$ e-mail: woodard@phys.ufl.edu
\end{flushleft}

\end{titlepage}

\section{Introduction}

Inflationary perturbations \cite{Starobinsky:1979ty,Mukhanov:1981xt} 
represent the first recognized quantum gravitational phenomena 
\cite{Woodard:2009ns,Ashoorioon:2012kh,Krauss:2013pha} and provide 
our most powerful tool for reconstructing the mechanism of primordial 
inflation \cite{Mukhanov:1990me,Liddle:1993fq,Lidsey:1995np}. These 
perturbations derive from 0-point fluctuations of gravitons and 
(in the simplest models) minimally coupled scalars on the 
background,
\begin{equation}
ds^2 = -dt^2 + a^2(t) d\vec{x} \!\cdot\! d\vec{x} \quad
\Longrightarrow \quad H(t) \equiv \frac{\dot{a}}{a} > 0 \quad ,
\quad \epsilon(t) \equiv -\frac{\dot{H}}{H^2} < 1 \; . \label{FLRW}
\end{equation}
The tensor and scalar mode functions, $u(t,k)$ and $v(t,k)$, are 
initially oscillating and red-shifting like those of normal particles 
\cite{Woodard:2014jba},
\begin{equation}
k \gg H(t) a(t) \quad \Longrightarrow \quad u(t,k) \simeq \frac{
\exp[-ik \!\int_{t_i}^{t} \!\! \frac{dt'}{a(t')}]}
{\sqrt{2 k a^2(t)}} \; , \; v(t,k) \simeq \frac{
\exp[-ik \!\int_{t_i}^{t} \!\! \frac{dt'}{a(t')}]}{\sqrt{2 k
\epsilon(t) a^2(t)}} \; . \label{UV} 
\end{equation}
However, after the time $t_k$ of first horizon crossing ($k =
H(t_k) a(t_k)$), one can see from their evolution equations,
\begin{equation}
\ddot{u} + 3 H \dot{u} + \frac{k^2}{a^2} u = 0 \qquad , \qquad
\ddot{v} + \Bigl( 3H \!+\! \frac{\dot{\epsilon}}{\epsilon}\Bigr)
\dot{v} + \frac{k^2}{a^2} v = 0 \; .
\end{equation}
that both mode functions approach constants of the form 
\cite{Brooker:2015iya,Brooker:2016xkx},
\begin{eqnarray}
k \ll H(t) a(t) & \Longrightarrow & \Bigl\vert u(t,k)\Bigr\vert^2
\longrightarrow \frac{H^2(t_k)}{2 k^3} \!\times\! C\Bigl( 
\epsilon(t_k)\Bigr) \!\times\! \mathcal{C}(k) \; , 
\label{IRu} \\
k \ll H(t) a(t) & \Longrightarrow & \Bigl\vert v(t,k)\Bigr\vert^2
\longrightarrow \frac{H^2(t_k)}{2 k^3 \epsilon(t_k)} \!\times\! 
C\Bigl( \epsilon(t_k)\Bigr) \!\times\! \mathcal{S}(k) \; .
\label{IRv}
\end{eqnarray}
Here the (monotonically decreasing) slow-roll correction factor is,
\begin{equation}
C(\epsilon) \equiv \frac1{\pi} \Gamma^2\Bigl( \frac12 \!+\! 
\frac1{1 \!-\! \epsilon}\Bigr) \Bigl[ 2 (1 \!-\! \epsilon)
\Bigr]^{\frac2{1-\epsilon}} \Longrightarrow \; C(0) = 1
\geq C(\epsilon) > C(1) = 0 \; ,
\end{equation}
while the nonlocal correction factors, $\mathcal{C}(k)$ and $\mathcal{S}(k)$,
are unity for $\dot{\epsilon} = 0$ and depend upon conditions only a few 
e-foldings before and after $t_k$. This transition from the ultraviolet, 
early-time form (\ref{UV}) to the infrared, late-time form 
(\ref{IRu}-\ref{IRv}) is known as {\it freezing-in}. It is how primordial 
tensor and scalar perturbations fossilize so that they can survive to the 
current epoch. 

The generation of perturbations is a tree order phenomenon but it has 
clear implications for loop corrections because the same tensor and scalar
mode functions appear in their respective propagators,
\begin{eqnarray}
i\Delta_{h}(x;x') & = \int \!\! \frac{d^3k}{(2\pi)^3}
\, e^{i\vec{k} \cdot (\vec{x} - \vec{x}')} \Biggl\{ \theta(t \!-\! t')
u(t,k) u^*(t',k) + \Bigl( t \leftrightarrow t'\Bigr) \Biggr\} , 
\label{Deltah} \\
i\Delta_{\zeta}(x;x') & = \int \!\! \frac{d^3k}{(2\pi)^3}
\, e^{i\vec{k} \cdot (\vec{x} - \vec{x}')} \Biggl\{ \theta(t \!-\! t')
v(t,k) v^*(t',k) + \Bigl( t \leftrightarrow t'\Bigr) \Biggr\} . 
\label{Deltaz}
\end{eqnarray}
Two related, and often confused, loop phenomena are {\it infrared divergences}
and {\it secular growth}. Infrared divergences arise because an infinite 
number of very long wavelength modes obey $k \ll H(t_i) a(t_i)$ even at the
time $t_i$ when inflation begins. Hence these modes begin life in the 
``saturated'' form (\ref{IRu}-\ref{IRv}). From the fact that $H(t)$ typically 
falls as inflation proceeds, and $\epsilon(t)$ typically grows, one can see 
that both $u(t,k) u^*(t',k)$ and $v(t,k) v^*(t',k)$ diverge more strongly than 
$1/k^3$ at $k = 0$, so the mode sums (\ref{Deltah}-\ref{Deltaz}) diverge. Note
that infrared divergence is due to the small $k$ behavior of the mode functions,
and is present even at the beginning of inflation. However, it would not occur
if one formulated inflation on a spatially closed manifold such as $T^3 \times 
R$ \cite{Tsamis:1993ub}, or if one assumed that the initially super-horizon 
modes were in some less infrared singular state than (\ref{IRu}-\ref{IRv}) 
\cite{Vilenkin:1983xp}. Infrared divergences were first noted in 1977, for 
$i\Delta_{h}(x;x')$ on constant $\epsilon(t)$ backgrounds on $R^3 \times R$, 
by Ford and Vilenkin \cite{Ford:1977in}.

In contrast, secular growth arises because the progression of inflation causes 
more and more initially ultraviolet modes to make the transition from the 
oscillatory ultraviolet form (\ref{UV}) to the saturated, infrared form 
(\ref{IRu}-\ref{IRv}). This endows the propagators (\ref{Deltah}-\ref{Deltaz})
with secular growth from the constructive interference of the ever-larger
number of super-horizon modes,
\begin{eqnarray}
i\Delta_{h}(x;x') \Bigl\vert_{\rm secular} & = & \frac1{4\pi^2} \! \int^{H a} 
\!\! \frac{dk}{k} H^2(t_k) \!\times\! C\Bigl(\epsilon(t_k)\Bigr)
\!\times\! \mathcal{C}(k) \; , \\
i\Delta_{\zeta}(x;x') \Bigl\vert_{\rm secular} & = & \frac1{4\pi^2} \! 
\int^{H a} \!\! \frac{dk}{k} \frac{H^2(t_k)}{\epsilon(t_k)} \!\times\! 
C\Bigl(\epsilon(t_k)\Bigr) \!\times\! \mathcal{S}(k) \; .
\end{eqnarray}
Unlike infrared divergences, secular growth occurs on spatially compact 
manifolds, and without regard to assumptions about initially super-horizon 
modes. It really has nothing to do with infrared divergences, except for
the fact that the late time form (\ref{IRu}-\ref{IRv}) happens to be the
same as the small $k$ limiting form. Secular growth of $i\Delta_{h}(x;x')$ 
was first noted in 1982, on de Sitter background ($\epsilon(t) = 0$), by 
Vilenkin and Ford \cite{Vilenkin:1982wt}, by Linde \cite{Linde:1982uu}, 
and by Starobinsky \cite{Starobinsky:1982ee},
\begin{equation}
\epsilon(t) = 0 \quad \Longrightarrow \quad i\Delta_{h}(x;x') 
\Bigl\vert_{\rm secular} = \frac{H^2}{4 \pi^2} \ln\Bigl( \sqrt{a(t) a(t')} 
\Bigr) \; .
\end{equation}
In 1987 Allen and Folacci demonstrated that the very same secular growth 
occurs on the full de Sitter manifold, which is spatially compact 
\cite{Allen:1987tz}.

The first proof that secular growth affects loop amplitudes came in 2002 
with a fully dimensionally regulated and renormalized evaluation of the 
expectation value of the stress tensor for a massless, minimally coupled
(MMC) scalar with a quartic self-interaction on a nondynamical de Sitter 
background \cite{Onemli:2002hr,Onemli:2004mb}. It was subsequently shown
that one and two loop corrections to the scalar mode functions of the
same theory also experience secular growth \cite{Brunier:2004sb,Kahya:2006hc}.
For (MMC) scalar quantum electrodynamics on nondynamical de Sitter background
secular growth has been seen in one loop corrections to the photon wave
function \cite{Prokopec:2002jn,Prokopec:2002uw}, in one loop corrections to 
electrodynamic forces \cite{Degueldre:2013hba}, as well as in the two loop 
expectation value of the stress tensor \cite{Prokopec:2008gw}. And secular
growth was demonstrated as well for Dirac fermions Yukawa-coupled to a MMC 
scalar on nondynamical de Sitter in the one loop correction to the fermion 
mode function \cite{Prokopec:2003qd,Garbrecht:2006jm}, and in the one loop 
expectation value of the stress tensor \cite{Miao:2006pn}. 

Each of these MMC scalar results can be understood using the stochastic 
formalism of Starobinsky \cite{Starobinsky:1986fx}, which has been proved to 
capture the leading secular effects of scalar potential models at each order 
in the loop expansion \cite{Tsamis:2005hd,Finelli:2008zg,Finelli:2010sh}. 
The Starobinsky formalism has been extended to scalar quantum electrodynamics 
\cite{Prokopec:2007ak} and Yukawa theory \cite{Miao:2006pn}. It even provides 
a nonperturbative resummation of the leading secular effects for those cases 
in which a time independent limit is approached \cite{Starobinsky:1994bd}.

Making quantum gravitational computations is vastly more difficult, but 
fully dimensionally regulated and BPHZ renormalized\footnote{The four initials
stand Bogoliubov, Parasiuk \cite{Bogoliubov:1957gp}, Hepp \cite{Hepp:1966eg}
and Zimmermann \cite{Zimmermann:1968mu,Zimmermann:1969jj}, who developed 
the standard technique of subtracting divergences with local counterterms,
even for nonrenormalizable theories like quantum gravity.} results have
been obtained on de Sitter background for one graviton loop corrections to 
MMC scalars, photons and fermions. MMC scalar mode functions experience
no secular corrections at one loop because the scalar only couples to the
metric through its kinetic energy, which red-shifts to zero \cite{Kahya:2007bc,
Kahya:2007cm}. In contrast, photons carry spin, which allows them to continue
interacting with inflationary gravitons to arbitrarily late times 
\cite{Leonard:2013xsa}. As a consequence, dynamical photons experience a
secular enhancement from inflationary gravitons \cite{Wang:2014tza}, as
do certain electrodynamic forces \cite{Glavan:2013jca}. The spin-spin
coupling between fermions and gravitons also gives rise to a persistent
interaction \cite{Miao:2008sp}. Hence inflationary gravitons also induce
a secular enhancement of fermions \cite{Miao:2005am,Miao:2006gj}.

No comparably explicit calculations have been performed on realistic 
inflationary backgrounds (which means $\dot{\epsilon} \neq 0$) because 
the mode functions and propagators are unknown. Even working out the 
interactions of the gauge-fixed and constrained theory has only been 
done to 3-point \cite{Maldacena:2002vr} and 4-point orders \cite{Seery:2006vu,
Jarnhus:2007ia,Xue:2012wi}. In spite of these limitations, an important 
theorem by Weinberg establishes that loops corrections to the power spectra 
can grow no faster than powers of the logarithm of the scale factor \cite{Weinberg:2005vy,Weinberg:2006ac}. There has also been a convincing
demonstration by Giddings and Sloth that infrared divergences from the 
graviton propagator (\ref{Deltah}) affect the inflationary power spectra at 
one loop order \cite{Giddings:2010nc}.

The technique of Giddings and Sloth has much to do with why so many
people concede the existence of secular effects from MMC scalars but
deny that they can occur from gravitons. It helps to change the temporal 
variable from $t$ to conformal time $\eta$ with $d\eta = dt/a(t)$, so 
that the background metric takes the form $a^2 \eta_{\mu\nu}$. Now
conformally transform the full metric by the scale factor and express 
the conformally transformed metric in terms of the graviton field 
$h_{\mu\nu}$,
\begin{equation}
g_{\mu\nu}(\eta,\vec{x}) \equiv a^2 \widetilde{g}_{\mu\nu}(\eta,\vec{x}) 
\equiv a^2 \Bigl[ \eta_{\mu\nu} + \kappa h_{\mu\nu}(\eta,\vec{x}) \Bigr]
\quad , \quad \kappa^2 \equiv 16 \pi G \; . \label{metric}
\end{equation}
At linearized order the graviton field can be expressed as a mode sum
over spatial plane waves and polarizations whose precise form is known
\cite{Tsamis:1992zt} but not relevant for our discussion,
\begin{equation}
h_{\mu\nu}(\eta,\vec{x}) = \int \!\! \frac{d^3k}{(2\pi)^3} \! 
\sum_{\lambda} \Biggl\{ u(\eta,k,\lambda) e^{i \vec{k} \cdot \vec{x}} 
\epsilon_{\mu\nu}(\vec{k},\lambda) \alpha(\vec{k},\lambda) + {\rm c.c.}
\Biggr\} . \label{modesum}
\end{equation}
Because the mode functions $u(\eta,k,\lambda)$ of the super-horizon 
($k < H(t) a(t)$) wavelengths freeze in to constant values, the 
super-horizon part of the mode sum behaves as a classical stochastic 
random field \cite{Vilenkin:1982wt,Starobinsky:1986fx,Rey:1986zk,
Sasaki:1987gy,Winitzki:1999ve}. This is still an operator by virtue of
the factors of $\alpha(\vec{k},\lambda)$ and 
$\alpha^{\dagger}(\vec{k},\lambda)$ but, if we neglect the very small 
residual time dependence of $u(\eta,k,\lambda)$, it commutes with its
time derivative and its value in any particular state does not 
change \cite{Tsamis:2010iw}. This means we can treat it as a constant. 
Of course a local observer would choose coordinates so as to absorb 
this constant. So the long wavelength modes have no effect in these
new coordinates --- because the long wavelength modes are not even 
present --- and their apparent effect in the original coordinates 
$x^{\mu} = (\eta,\vec{x})$ is just the result of evaluating tree order 
results at the transformed coordinates.

This nice insight by Giddings and Sloth \cite{Giddings:2010nc,
Giddings:2010ui}, which was anticipated by Urakawa and Tanaka 
\cite{Urakawa:2009my,Urakawa:2009gb,Urakawa:2010it,Urakawa:2010kr}, 
is valid for the case of infrared divergences because the modes which 
cause them are in the saturated state from the beginning of inflation. 
It is not clear that one can apply the same insight to the case of 
secular dependence because the modes responsible for that were initially
sub-horizon, with nontrivial spacetime dependence, and they only later
experienced freeze-in. An example of the potential problems was given
in equations (41-43) and the associated discussion of \cite{Miao:2012xc}. 
However, enthusiasm over progress on the very the tough problem of 
computing loop corrections to the power spectra made it inevitable that 
such applications would appear \cite{Giddings:2011zd,Giddings:2011ze,
Tanaka:2011aj,Urakawa:2011fg,Tanaka:2012wi,Tanaka:2013xe,Tanaka:2013caa,
Tanaka:2014ina}. Related, and perhaps contradictory, claims have also 
been made that the power spectra of single scalar inflation show no 
secular corrections at all once changes in the perturbative background 
are properly incorporated \cite{Senatore:2012nq,Senatore:2012wy,
Pimentel:2012tw}.

The various authors who deny the existence of secular corrections from 
inflationary gravitons are focussed narrowly on the special case of the 
power spectra for single-scalar inflation. However, there seems nothing 
about the key argument which restricts its applicability, either as 
regards the model of inflation or the quantity under study. We shall 
therefore formalize their belief in the {\it Transformation Ansatz}: 
that secular loop corrections to any quantity from inflationary 
gravitons are the result of evaluating the tree order quantity at the 
transformed coordinates which would render the metric (\ref{metric}) 
conformal to $\eta_{\mu\nu}$ for an exactly constant $\widetilde{g}_{
\mu\nu}(\eta,\vec{x})$.

The purpose of this paper is to test the transformation ansatz by
working out its consequences for photons and fermions at one loop
order on de Sitter background, and then comparing with the exact
results of dimensionally regulated and renormalized computations
\cite{Leonard:2013xsa,Wang:2014tza,Miao:2005am,Miao:2006gj}. If 
the ansatz is correct then the leading secular effects will agree.
In section 2 we work out the coordinate transformation and the 
associated, local Lorentz transformation, which would carry a truly
constant $\widetilde{g}_{\mu\nu}$ back to $\eta_{\mu\nu}$. In section
3 we apply this transformation to the free photon wave function, and
then compute the leading secular corrections. Section 4 does the same
thing for the free fermion mode function. Our conclusions comprise
section 5. 

\section{Constructing the Transformations}

The purpose of this section is to construct the general coordinate 
transformation, and the associated local Lorentz transformation, which 
would carry the metric $g_{\mu\nu} = a^2 \widetilde{g}_{\mu\nu}$ to the 
pure de Sitter form $g'_{\mu\nu} = a^2 \eta_{\mu\nu}$ under the (false) 
assumption that $\widetilde{g}_{\mu\nu}$ is a spacetime constant. We
begin by giving the vierbein in Lorentz-symmetric gauge, which plays a 
prominent role in the construction. Then the general coordinate 
transformation is derived. The section closes by working out the 
Dirac spinor Lorentz transformation this general coordinate transformation
induces. To simplify the discussion we commit a small abuse of our earlier
notation by considering the de Sitter scale factor to be a function of the 
conformal time $a(\eta) \equiv -1/H\eta$.

\subsection{The Vierbein in Symmetric Gauge}

When considering theories with half-integral spin coupled to gravity it is 
convenient to introduce a fictitious local Lorentz gauge symmetry under 
which the spinor indices transform. The place of the metric is taken by the 
vierbein $e_{\mu a}(x)$, with vector index $\mu$ and local Lorentz index $a$. 
The metric follows by Lorentz contracting two vierbeins, $g_{\mu\nu}(x) = 
e_{\mu a}(x) e_{\nu b}(x) \eta^{ab}$. For our conformally transformed metric $\widetilde{g}_{\mu\nu}$ the associated vierbein would be 
$\widetilde{e}_{\mu a}$,
\begin{equation}
\widetilde{g}_{\mu\nu}(x) = \widetilde{e}_{\mu a}(x) \widetilde{e}_{\nu b}(x)
\eta^{ab} \quad , \quad \widetilde{e}^{\mu}_{~a} \widetilde{e}_{\mu b} =
\eta_{ab} \; . 
\end{equation}
The fictitious nature of local Lorentz symmetry is evidenced by its failure
to obey the famous rule of van Nieuwenhuizen (for real symmetries) that 
``gauge fixing counts twice.'' Once local Lorentz symmetry has been gauge 
fixed, the associated constraint equations are automatically obeyed, instead 
of imposing nontrivial relations between the surviving fields. That is how 
the 16 components of the vierbein reduce to the usual two graviton degrees 
of freedom,
\begin{eqnarray}
\lefteqn{\Bigl( {\rm 2\ gravitons}\Bigr) = \Bigl({\rm 16\ fields}\Bigr) - 
\Bigl({\rm 4\ coordinate\ gauges}\Bigr) - \Bigl({\rm 6\ Lorentz\ gauges}\Bigr) 
} \nonumber \\
& & \hspace{2cm} - \Bigl({\rm 4\ coordinate\ constraints}\Bigr)
 - \Bigl({\rm 0\ Lorentz\ constraints}\Bigr) \; . \qquad
\end{eqnarray}  
Had local Lorentz symmetry been real, the counting would have produced the 
absurd result $16 - 4 - 6 - 4 - 6 = -4$! Giving trivial constraints is one way 
to recognize when compensating fields have been introduced to make a theory
appear to possesses some symmetry it really does not have \cite{Tsamis:1984hh}.

Although local Lorentz invariance is fictitious, its gauge fixing still
induces a Faddeev-Popov determinant, which can complicate calculations.
{\it Lorentz symmetric gauge} ($\widetilde{e}_{\mu a} = \widetilde{e}_{a \mu}$)
is a particularly nice condition for which the Faddeev-Popov determinant is unity
\cite{Woodard:1984sj}. Any local Lorentz gauge allows one to solve for the
vierbein in terms of the metric. For Lorentz symmetric gauge with
$\widetilde{g}_{\mu\nu} = \eta_{\mu\nu} + \kappa h_{\mu\nu}$ this solution
is \cite{Woodard:1984sj},
\begin{equation}
\widetilde{e}_{\mu a} = \widetilde{e}_{a \mu} = \Bigl( 
\sqrt{\widetilde{g} \, \eta^{-1}} \, \Bigr)_{\mu}^{~\nu} \!\times\! 
\eta_{\nu a} = \eta_{\mu a} + \frac12 \kappa h_{\mu a} - \frac18 
\kappa^2 h_{\mu}^{~\nu} h_{\nu a} + \dots \label{vierbein}
\end{equation}
The inverse vierbein is,
\begin{equation}
\widetilde{e}^{\mu}_{~a} \equiv \widetilde{g}^{\mu\nu} 
\widetilde{e}_{\nu a} = \delta^{\mu}_{~a} - \frac12 \kappa h^{\mu}_{~a}
+ \frac38 \kappa^2 h^{\mu\nu} h_{\nu a} - \dots \label{invvier}
\end{equation}

\subsection{The General Coordinate Transformation}

A general coordinate transformation $x^{\mu} \rightarrow {x'}^{\mu}(x)$
carries the metric to,
\begin{equation}
g'_{\mu\nu}(x') = \frac{\partial x^{\rho}(x')}{\partial {x'}^{\mu}}
\frac{\partial x^{\sigma}(x')}{\partial {x'}^{\nu}} g_{\rho\sigma}(x)
\;\; \Longleftrightarrow \;\; g'_{\mu\nu}(x) = 
\frac{\partial x^{\rho}(x)}{\partial {x'}^{\mu}}
\frac{\partial x^{\sigma}(x)}{\partial {x'}^{\nu}} 
g_{\rho\sigma}\Bigl({x'}^{-1}\Bigr) \; .
\end{equation}
Under the (false) assumption that $\widetilde{g}_{\mu\nu}$ is constant
in space and time it is clear that we seek a linear transformation,
\begin{equation}
{x'}^{\mu} = \Omega^{\mu}_{~\nu} x^{\nu} \quad \Longleftrightarrow \quad
x^{\mu} = \omega^{\mu}_{~\nu} {x'}^{\nu} \qquad , \qquad 
\Omega^{\mu}_{~\rho} \omega^{\rho}_{~\nu} = \delta^{\mu}_{~\nu} =
\omega^{\mu}_{~\rho} \Omega^{\rho}_{~\nu} \; .
\end{equation}
We require that the transformation makes the metric conformal,
\begin{equation}
g'_{\mu\nu}(x) = \omega^{\rho}_{~\mu} \omega^{\sigma}_{~\nu} 
g_{\rho\sigma}(\omega x) = a^2\Bigl(\omega^0_{~\rho} x^{\rho}\Bigr) 
\eta_{\mu\nu} \qquad \Longrightarrow \qquad 
\widetilde{g}_{\rho\sigma} \omega^{\rho}_{~\mu} \omega^{\sigma}_{~\nu} = 
\eta_{\mu\nu} \; , \label{condition1}
\end{equation} 
and also that the proportionality factor depends only on conformal time, 
although it may have a different Hubble constant,
\begin{equation}
\omega^{0}_{~\mu} = \frac{H'}{H} \!\times\! \delta^{0}_{~\mu} 
\; . \label{condition2} 
\end{equation}

The first condition (\ref{condition1}) is achieved by the symmetric 
gauge vierbein,
\begin{equation}
\widetilde{g}_{\rho\sigma} \widetilde{e}^{\rho}_{~a} 
\widetilde{e}^{\sigma}_{~b} = \eta_{ab} \ ; \label{ok1}
\end{equation}
However, the vierbein does not generally obey the condition (\ref{condition2}). 
To enforce this without disturbing (\ref{condition1}) we concatenate a Lorentz 
transformation,
\begin{equation}
\omega^{\mu}_{~\nu} = \widetilde{e}^{\mu}_{~a} \Lambda^{a}_{~\nu} \; .
\label{diffeo}
\end{equation}
The desired transformation takes the form of a boost whose $3+1$ decomposition 
can be expressed in terms of a 3-velocity $\beta^i = \beta \widehat{\beta}^i$ 
with $\gamma = 1/\sqrt{1 - \beta^2}$,\footnote{We follow the usual $3+1$ 
convention of making no distinction between upper and lower indices for
intrinsically spatial quantities such as $\widehat{\beta}^{n} = 
\widehat{\beta}_{n}$ and $\delta^{mn} = \delta^{m}_{~n}$.} 
\begin{equation}
\Lambda^{\mu}_{~\nu} \equiv \left( \matrix{ \Lambda^{0}_{~0} & 
\Lambda^{0}_{~n} \cr \Lambda^{m}_{~0} & \Lambda^{m}_{~n} } \right) =
\left( \matrix{ \gamma & - \beta \gamma \widehat{\beta}^{n} \cr
-\beta \gamma \widehat{\beta}^{m} & \delta^{mn} + (\gamma \!-\! 1)
\widehat{\beta}^{m} \widehat{\beta}^{n} }\right) \; . \label{Lambda3+1}
\end{equation}
The $3+1$ expression for the full transformation is,
\begin{eqnarray}
\left( \matrix{\omega^{0}_{~0} & \omega^{0}_{~n} \cr
\omega^{m}_{~0} & \omega^{m}_{~n} } \right) & = & 
\left( \matrix{\widetilde{e}^{0}_{~0} & \widetilde{e}^{0}_{~i} \cr
\widetilde{e}^{m}_{~0} & \widetilde{e}^{m}_{~i} } \right) \times
\left( \matrix{\Lambda^{0}_{~ 0} & \Lambda^{0}_{~n} \cr
\Lambda^{i}_{~0} & \Lambda^{i}_{~n} } \right) \; , \\
& = & \left( \matrix{\widetilde{e}^{0}_{~0} \Lambda^{0}_{~0} \!+\! 
\widetilde{e}^{0}_{~i} \Lambda^{i}_{~0} & 
\widetilde{e}^{0}_{~0} \Lambda^{0}_{~n} \!+\! 
\widetilde{e}^{0}_{~i} \Lambda^{i}_{~n} \cr
\widetilde{e}^{m}_{~0} \Lambda^{0}_{~0} \!+\! 
\widetilde{e}^{m}_{~i} \Lambda^{i}_{~0} & 
\widetilde{e}^{m}_{~0} \Lambda^{0}_{~n} \!+\! 
\widetilde{e}^{m}_{~i} \Lambda^{i}_{~n} } \right) \; .
\end{eqnarray}
Condition (\ref{condition2}) requires $\omega^{0}_{~n} = 0$, which implies,
\begin{equation}
\widehat{\beta}^{i} = \frac{\widetilde{e}^{0}_{~i}}{\sqrt{ 
\widetilde{e}^{0}_{~j} \widetilde{e}^{0}_{~j}}} \quad , \quad
\beta^{i} = \frac{\widetilde{e}^{0}_{~i}}{\widetilde{e}^{0}_{~0}}
\quad , \quad \gamma = \frac{\widetilde{e}^{0}_{~0}}{\sqrt{-
\widetilde{g}^{00}}} \; . \label{betagamma}
\end{equation}
The time-time component of the transformation gives us the multiplicative 
change in the Hubble constant,\footnote{It is amusing to note that the 
order $\kappa^4$ contribution implies secular back-reaction at two loop 
order, which is something else which the sceptics disbelieve 
\cite{Garriga:2007zk,Tsamis:2008zz}.} 
\begin{equation}
\frac{H'}{H} = \omega^{0}_{~0} = \sqrt{-\widetilde{g}^{00}} = 1 + \frac12
\kappa h_{00} + \frac38 \kappa^2 h_{00}^2 - \frac12 \kappa^2 h_{0i} h_{0i} 
+ \dots \label{omega00}
\end{equation}
The space-time component can be expressed in terms of the metric,
\begin{equation}
\omega^{m}_{~0} = -\frac{\widetilde{g}^{0 m}}{\sqrt{-\widetilde{g}^{00}}} 
= -\kappa h_{m0} - \frac12 \kappa^2 h_{00} h_{0m} + \kappa^2 h_{0i} h_{im}
+ \dots \label{omegam0}
\end{equation}
The space-space component has a superficially complicated form that can be 
recognized as the inverse of the 3-dimensional driebein,
\begin{eqnarray}
\omega^{m}_{~n} & = & \widetilde{e}^{m}_{~n} - \frac{\widetilde{e}^{m}_{~0}
\widetilde{e}^{0}_{~n}}{\sqrt{-\widetilde{g}^{00}}} + 
\frac{ \widetilde{e}^{m}_{~i} \widetilde{e}^{0}_{~i} \widetilde{e}^{0}_{~n}}{
\widetilde{e}^{0}_{~j} \widetilde{e}^{0}_{~j}} 
\Bigl[ \frac{\widetilde{e}^{0}_{~0}}{\sqrt{-\widetilde{g}^{00}}} - 1 \Bigr]
\; , \\
& = & \delta_{mn} - \frac12 \kappa h_{mn} + \frac38 \kappa^2 h_{mi} h_{ni} 
+ \dots \label{omegamn}
\end{eqnarray}

\subsection{The Associated Dirac Spinor Transformation}

We stress that local Lorentz symmetry is completely fictitious. When one
fixes it by imposing some local Lorentz gauge condition then a general
coordinate transformation will generally disrupt the condition. One defines
the associated local Lorentz transformation by requiring it to restore the 
gauge condition. This is how spinor indices are transformed in general 
relativity.

Using expression (\ref{diffeo}) one can see that our general coordinate 
transformation takes the vierbein to,
\begin{equation}
\widetilde{e}'_{\mu a} = \omega^{\rho}_{~\mu} \widetilde{e}_{\rho a} =
\widetilde{e}^{\rho}_{~b} \Lambda^{b}_{~ \mu} \widetilde{e}_{\rho a}
= \eta_{ab} \Lambda^{b}_{~\mu} \; . \label{vierdif}
\end{equation}
From the Lorentz invariance of $\eta_{ab}$ we see that the additional, local
Lorentz transformation, needed to restore $\widetilde{e}'_{\mu a}$ to 
symmetric gauge is the very same one (\ref{Lambda3+1}) with parameters
(\ref{betagamma}),
\begin{equation}
\widetilde{e}'_{\mu a} \Lambda^{a}_{~c} = \eta_{ab} \Lambda^{b}_{~\mu} 
\Lambda^{a}_{~c} = \eta_{\mu c} \; .
\end{equation}

It remains to construct the Dirac spinor representation which corresponds to 
the boost (\ref{betagamma}). The general spinor Lorentz transformation takes
the form,
\begin{equation}
\Lambda_{ij} = \exp\Bigl[-\frac{i}{2} \theta_{ab} \mathcal{J}^{ab}\Bigr]_{ij}
\qquad , \qquad \Bigl[\mathcal{J}\Bigr]^{ab} = \frac{i}{4} \Bigl[\gamma^{a},
\gamma^{b}\Bigr]_{ij} \; , 
\end{equation}
where the gamma matrices are $\gamma^{a}_{~ij}$. The boost (\ref{betagamma})
is achieved by choosing the infinitesimal parameters as,
\begin{equation}
\theta^{0j} = \widehat{\beta}^{j} {\rm tanh}^{-1}(\beta) \qquad , \qquad 
\theta^{ij} = 0 \; .
\end{equation}
The spinor transformation takes the 2-component form,
\begin{equation}
\Lambda = \left( \matrix{ B & 0 \cr 0 & B^{-1}} \right) \qquad , \qquad B = 
\sqrt{\frac12 (1 \!+\! \gamma)} \, I + \frac{\gamma \vec{\beta} \!\cdot\!
\vec{\sigma}}{\sqrt{2 (1 \!+\! \gamma)}} \; , \label{LAMBDA}
\end{equation}  
where $I$ is the $2 \times 2$ unit matrix and $\vec{\sigma} \equiv (\sigma_1 ,
\sigma_2 , \sigma_3)$ are the Pauli matrices. Expanding the $2 \times 2$ 
matrix $B$ gives,
\begin{equation}
B = \Biggl[ 1 \!+\! \frac{9 \kappa^2}{32} h_{00}^2 \!-\! \frac{\kappa^2}4 
h_{0i} h_{0i} \!+\! \dots\Biggr] I + \Biggl[ \frac{\kappa}4 h_{0i} \!+\! 
\frac{\kappa^2}{16} h_{00} h_{0i} \!-\! \frac{3\kappa^2}8 h_{ij} h_{0j} \!+\! 
\dots\Biggr] \sigma_i \; . \label{Bexp}
\end{equation}

\section{The Photon Polarization Vector}

The purpose of this section is to compare exact one loop results with the
predictions of the transformation ansatz for the leading secular corrections 
to the photon polarization vectors from quantum gravity at one loop 
($\kappa^2$) order. We begin by summarizing the exact one loop computation
\cite{Leonard:2013xsa,Wang:2014tza}. Then we use the results of the previous 
section to derive the prediction of the transformation ansatz.

\begin{figure}[ht]
\includegraphics[width=13cm,height=5cm]{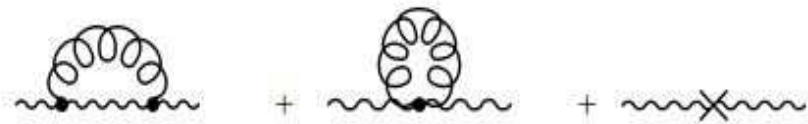}
\caption{Feynman diagrams which comprise the one loop quantum
gravitaitonal contribution to the vacuum polarization. Wavy lines
stand for photons; wiggly lines stand for gravtions.}
\label{PhotonDiagrams}
\end{figure}

\subsection{The Exact Computation}

The computation was performed in two steps. First, dimensional regularization 
and BPHZ renormalization were employed to evaluate the diagrams of 
Figure~\ref{PhotonDiagrams} giving the one graviton loop contribution
to the vacuum polarization $-i[\mbox{}^{\mu} \Pi^{\nu}](x;x')$ on de Sitter 
background \cite{Leonard:2013xsa}. In the second step, the linearized 
Schwinger-Keldysh effective field equations were solved for plane wave 
photons $A_{\mu}(x) = \epsilon_{\mu}(\eta,\vec{k},\lambda) \times e^{i \vec{k} 
\cdot \vec{x}}$ \cite{Wang:2014tza},
\begin{equation}
\partial_{\nu} \Bigl[\sqrt{-g} \, g^{\nu\rho} g^{\mu\sigma} F_{\rho\sigma}(x)
\Bigr] + \int \!\! d^4x' \Bigl[ \mbox{}^{\mu} \Pi^{\nu}\Bigr](x;x') A_{\nu}(x')
= 0 \; . \label{Maxeqn}
\end{equation}

Because 4-dimensional photons are conformally invariant, the tree order 
polarization vector on de Sitter is identical (in conformal coordinates) to 
the usual flat space one,
\begin{equation}
\epsilon^{(0)}_{\mu}(\eta,\vec{k},\lambda) = \frac{e^{-ik \eta}}{\sqrt{2k}}
\left( \matrix{ 0 \cr \frac1{\sqrt{2}} [\widehat{\theta} \!+\! i \lambda
\widehat{\phi}] }\right) \equiv \left( \matrix{\epsilon^{(0)}_{0}(
\eta,\vec{k},\lambda) \cr \epsilon^{(0)}_{i}(\eta,\vec{k},\lambda)} \right)
\; , \label{treephoton}
\end{equation}
where $\widehat{\theta}$ and $\widehat{\phi}$ are the other orthogonal unit
vectors in the system where the momentum $\vec{k} = k \widehat{r}$ is radial. 
At late times the one loop correction takes the form \cite{Wang:2014tza},
\begin{equation}
\epsilon^{(1)}_{\mu}(\eta,\vec{k},\lambda) \longrightarrow 
\frac{\kappa^2 H^2}{8 \pi^2} \, \frac{i k \ln(a)}{H a} \!\times\! 
\epsilon^{(0)}_{\mu}(\eta,\vec{k},\lambda) \; . \label{photonresult}
\end{equation}
Although the one loop polarization vector (\ref{photonresult}) actually falls 
off with respect to tree order result (\ref{treephoton}), its approach to
zero is slower than the latter's approach to a constant. Hence the one loop 
electric field strength grows relative to the tree result \cite{Wang:2014tza},
\begin{eqnarray}
F^{(1)}_{0i}(\eta,\vec{x}) & \longrightarrow & \frac{\kappa^2 H^2}{8 \pi^2} \, 
\ln(a) \!\times\! F^{(0)}_{0i}(\eta,\vec{x}) \; , \label{photongrowth} \\
F^{(1)}_{ij}(\eta,\vec{x}) & \longrightarrow & \frac{\kappa^2 H^2}{8 \pi^2} \, 
\frac{ik \ln(a)}{H a} \!\times\! F^{(0)}_{ij}(\eta,\vec{x}) \; .
\end{eqnarray}
This growth must eventually lead to a breakdown of perturbation theory. The 
physical interpretation of (\ref{photongrowth}) seems to be that the photon's 
physical 3-momentum redshifts like $k/a$, so 3-momentum tends to be added by 
scattering with the ensemble of inflationary gravitons, whose peak 3-momenta 
remains at about $H$ due to continual production.

\subsection{Prediction of the Transformation Ansatz}

The graviton propagator in the gauge which was used for the explicit 
computations \cite{Leonard:2013xsa,Wang:2014tza,Miao:2005am,Miao:2006gj}
takes the form of a sum of scalar propagators times constant tensors
\cite{Tsamis:1992xa,Woodard:2004ut},
\begin{equation}
i\Bigl[ \mbox{}_{\mu\nu} \Delta_{\rho\sigma}\Bigr](x;x') = 
\sum_{I = A,B,C} i\Delta_{I}(x;x') \!\times\! \Bigl[ \mbox{}_{\mu\nu}
T_{\rho\sigma}\Bigr] \; .
\end{equation}
The ($D$-dimensional) tensor factors are expressed in terms of the
purely spatial Lorentz metric $\overline{\eta}_{\mu\nu} \equiv 
\eta_{\mu\nu} + \delta^{0}_{\mu} \delta^{0}_{\nu}$,
\begin{eqnarray}
\Bigl[ \mbox{}_{\mu\nu} T^A_{\rho\sigma}\Bigr] & = & 
\overline{\eta}_{\mu \rho} \overline{\eta}_{\sigma \nu} +
\overline{\eta}_{\mu \sigma} \overline{\eta}_{\rho \nu} -
\frac{2}{D \!-\! 3} \, \overline{\eta}_{\mu\nu} 
\overline{\eta}_{\rho\sigma} \; , \\
\Bigl[ \mbox{}_{\mu\nu} T^B_{\rho\sigma}\Bigr] & = & - 
\delta^{0}_{\mu} \overline{\eta}_{\nu \rho} \delta^{0}_{\sigma} \!-\!
\delta^{0}_{\mu} \overline{\eta}_{\nu \sigma} \delta^{0}_{\rho} \!-\!
\delta^{0}_{\nu} \overline{\eta}_{\mu \rho} \delta^{0}_{\sigma} \!-\!
\delta^{0}_{\nu} \overline{\eta}_{\mu \sigma} \delta^{0}_{\rho} \; , \\
\Bigl[ \mbox{}_{\mu\nu} T^C_{\rho\sigma}\Bigr] & = & 
\frac{2}{(D \!-\! 2) (D \!-\! 3)} \Bigl[(D \!-\! 3) \delta^{0}_{\mu} 
\delta^{0}_{\nu} \!+\! \overline{\eta}_{\mu \nu} \Bigr] \Bigl[
(D \!-\! 3) \delta^{0}_{\rho} \delta^{0}_{\sigma} \!+\! 
\overline{\eta}_{\rho \sigma} \Bigr] \; .
\end{eqnarray}
The $A$-type, $B$-type and $C$-type propagators are those of minimally
coupled scalars with masses,
\begin{equation}
m^2_{A} = 0 \quad ,\quad m^2_{B} = (D \!-\! 2) H^2 \quad , \quad
m^2_{C} = 2 (D \!-\! 3) H^2 \; .
\end{equation}
Their full spacetime dependence is well known 
\cite{Tsamis:1992xa,Woodard:2004ut} but the only thing we require for
this analysis is their coincidence limits,
\begin{equation}
i\Delta_{A}(x;x) = {\rm Constant} + \frac{H^2}{4 \pi^2} \ln(a) 
\quad , \quad i\Delta_{B}(x;x) = {\rm Constant} = i\Delta_{C}(x;x) \; .
\end{equation}
The leading secular growth comes from just the logarithm in 
$i\Delta_A(x;x)$ which, by the way, agrees with the results of Vilenkin 
and Ford \cite{Vilenkin:1982wt}, Linde \cite{Linde:1982uu},
Starobinsky \cite{Starobinsky:1982ee} and Allen and Folacci 
\cite{Allen:1987tz}. For our purposes the expectation value of two
gravitons is therefore,
\begin{equation}
\Bigl\langle \Omega \Bigl\vert \kappa h_{\mu\nu}(x) \kappa h_{\rho\sigma}(x) 
\Bigr\vert \Omega \Bigr\rangle \longrightarrow \frac{\kappa^2 H^2 \ln(a)}{
4 \pi^2} \!\times\! \Bigl[ \overline{\eta}_{\mu \rho} 
\overline{\eta}_{\nu\sigma} \!+\! \overline{\eta}_{\mu\sigma} 
\overline{\eta}_{\nu\rho} \!-\! 2 \overline{\eta}_{\mu\nu} 
\overline{\eta}_{\rho\sigma} \Bigr] \; . \label{VEV}
\end{equation}

The transformation ansatz asserts that the leading secular growth in
the solution of (\ref{Maxeqn}) comes from taking the expectation value,
in the graviton vacuum, of the transformed tree order solution,
\begin{equation}
A^{\rm TA}_{\mu}(x) \equiv \Bigl\langle \Omega \Bigl\vert
\omega^{\nu}_{~\mu} \epsilon^{(0)}_{\nu}\Bigl( \omega^{0}_{~\rho} x^{\rho},
\vec{k},\lambda\Bigr) e^{i k_{j} \omega^{j}_{~\sigma} x^{\sigma}} 
\Bigr\vert \Omega \Bigr\rangle \; . \label{photonansatz}
\end{equation}
The matrix $\omega^{\mu}_{~\nu}$ is the one we constructed in section 2.2, 
with the graviton fields evaluated at $x^{\mu} = (\eta,\vec{x})$. Of course 
the transformation derived in section 2.2 was only valid for spacetime {\it 
constant} graviton fields. When the spacetime dependence of the graviton 
field becomes significant there is no transformation which can re-impose
de Sitter background. We therefore expect the transformation ansatz to
disagree with the exact computation (\ref{photonresult}), but it is worth
making the comparison in order to demonstrate that secular growth is
neither the same as infrared divergence, nor is it pure gauge.

Expression (\ref{VEV}) shows that only the purely
spatial components of the graviton field contribute secular growth
factors. We can therefore drop the temporal components from series 
expansions (\ref{omega00}), (\ref{omegam0}) and (\ref{omegamn}),
\begin{equation}
\omega^{\mu}_{~\nu} \equiv
\left( \matrix{\omega^{0}_{~0} & \omega^{0}_{~n} \cr
\omega^{m}_{~0} & \omega^{m}_{~n}} \right) \longrightarrow
\left( \matrix{ 0 & 0 \cr 0 & \delta_{mn} \!-\! \frac{\kappa}{2} h_{mn} 
+ \frac{3 \kappa^2}{8} h_{m i} h_{ni} \!+\! \dots} \right) \; . 
\label{omeganotime}
\end{equation}
There are two factors in (\ref{photonansatz}), the spatial plane wave,
\begin{equation}
e^{i k_j \omega^{j}_{~\sigma} x^{\sigma}} = e^{i \vec{k} \cdot \vec{x}}
\Biggl\{ 1 \!-\! \frac{i \kappa}{2} h_{mn} k^{m} x^{n} \!+\! 
\frac{i 3 \kappa^2}{8} h_{m\ell} h_{n\ell} k^{m} x^{n} \!-\! 
\frac{\kappa^2}{4} \Bigl( h_{mn} k^{m} k^{n}\Bigr)^2 \!+\! \dots
\Biggr\} \, \label{planewave} 
\end{equation}
and the (purely spatial) vector transformation,
\begin{equation}
\omega^{j}_{~ i} \epsilon^{(0)}_{j} = \epsilon^{(0)}_{i} \!-\! 
\frac{\kappa}{2} h_{ij} \epsilon^{(0)}_{j} \!+\! \frac{3 \kappa^2}{8}
h_{i k} h_{jk} \epsilon^{(0)}_{j} \!+\! \dots \label{vector}
\end{equation}
Multiplying (\ref{vector}) by (\ref{planewave}) and taking the expectation 
value using (\ref{VEV}) gives,
\begin{eqnarray}
\lefteqn{A^{\rm TA}_{i}(x) = \epsilon^{(0)}_{i}(\eta,\vec{k},\lambda) 
e^{i \vec{k} \cdot \vec{x}} + } \nonumber \\
& & \hspace{-.7cm} \frac{ \kappa^2 H^2 \ln(a)}{16 \pi^2} \Biggl[\!\Bigl(3 \!+\!
i\vec{k} \!\cdot\! \vec{x} \!-\! k^2 x^2 \!+\! (\vec{k} \!\cdot\! \vec{x})^2
\Bigr) \delta_{ij} \!+\! i k_{i} x_{j}\Biggr] \epsilon^{(0)}_{j}(\eta,\vec{k},
\lambda) e^{i \vec{k} \cdot \vec{x}} \!\!+\! O(\kappa^4) . \qquad \label{TAphoton}
\end{eqnarray}
As expected, the prediction (\ref{TAphoton}) of the transformation ansatz 
disagrees with the exact one loop computation (\ref{photonresult}). Not only is
(\ref{TAphoton}) stronger by a (huge) scale factor, it also contains some strange
factors of the 3-momentum and the spatial position. 

\section{The Fermion Wave Function}

The purpose of this section is to compare exact one loop results with the
predictions of the transformation ansatz for the leading secular corrections 
to the fermion mode functions from quantum gravity at one loop ($\kappa^2$) 
order. We begin by summarizing the exact one loop computation
\cite{Miao:2005am,Miao:2006gj}. Then we use the results of the previous 
section to derive the prediction of the transformation ansatz.

\begin{figure}[ht]
\includegraphics[width=13cm,height=5cm]{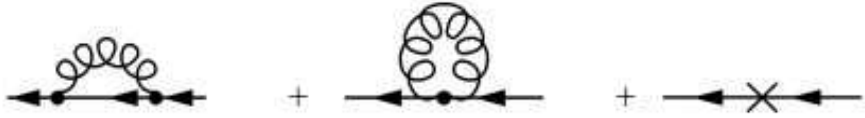}
\caption{Feynman diagrams which comprise the one loop quantum
gravitational contribution to the fermion self-energy. Solid
lines stand for fermions; wiggly lines stand for gravitons.}
\label{FermionDiagrams}
\end{figure}

\subsection{The Exact Computation}

Like the electromagnetic analogue (which was actually performed afterwards) 
the computation was made in two steps. First, dimensional regularization 
and BPHZ renormalization were employed to evaluate the diagrams of 
Figure~\ref{FermionDiagrams} to give the one graviton loop contribution
to the self-energy $-i[\mbox{}_{i} \Sigma_{j}](x;x')$ of massless Dirac 
fermions on de Sitter background \cite{Miao:2005am}. In the second step,
the linearized Schwinger-Keldysh effective field equations were solved
for (conformally rescaled) plane wave fermions $\Psi_i(x) = 
u_i(\eta,\vec{k},s) \times e^{i \vec{k} \cdot \vec{x}}$ \cite{Miao:2006gj},
\begin{equation}
i \gamma^{\mu}_{ij} \partial_{\mu} \Psi_{j}(x) - \int \!\! d^4x' 
\Bigl[ \mbox{}_{i} \Sigma_{j}\Bigr](x;x') \Psi_{j}(x') = 0 \; .
\end{equation}

Because massless fermions are conformally invariant, the tree order mode 
function on de Sitter is identical to the usual flat space one,
\begin{equation}
u^{(0)}_{i}(\eta,\vec{k},s) = \frac{e^{-ik \eta}}{\sqrt{2}} 
\left( \matrix{[I \!-\! \widehat{k} \!\cdot\! \vec{\sigma}] \xi(s) \cr
[I \!+\! \widehat{k} \!\cdot\! \vec{\sigma}] \xi(s)} \right) \; ,
\end{equation} 
where $I$ is the $2 \times 2$ unit matrix, $\vec{\sigma} = (\sigma_1, 
\sigma_2, \sigma_3)$ are the Pauli matrices and $\xi(s)$ is a 2-component
spinor. At late times the one loop correction takes the form 
\cite{Miao:2006gj},
\begin{equation}
u^{(1)}_{i}(\eta,\vec{k},s) \longrightarrow \frac{17}{2} 
\frac{\kappa^2 H^2}{2^6 \pi^2} \, \ln(a) \!\times\! 
u^{(0)}_{i}(\eta,\vec{k},s) \; . \label{fermionresult}
\end{equation}
The physical interpretation seems to be that the ensemble of
inflationary gravitons which pervades space scatters the propagating 
fermion by an amount which grows larger the farther in space (and hence 
the longer in time) the fermion propagates. One fascinating consequence 
of the secular growth evident in (\ref{fermionresult}) is that 
perturbation theory must eventually break down.

A vertex-by-vertex examination of the computation revealed that the
secular growth factors of $\ln(a)$ in (\ref{fermionresult}) all derive 
from the singly differentiated graviton fields of the spin connection 
\cite{Miao:2008sp}. This is consistent with the physical insight that 
the spin-spin coupling between fermions and gravitons allows even a 
highly redshifted fermion to continue interacting with inflationary 
gravitons. That makes good sense but it is completely at odds with the 
basic assumption of the transformation ansatz that there is no distinction 
between infrared divergences and secular effects, and that spacetime 
constant field configurations (which would be annihilated by differentiation) 
are responsible for both phenomena. We therefore expect that the
transformation ansatz will fail to recover the result (\ref{fermionresult}) 
of explicit computation.

\subsection{Prediction of the Transformation Ansatz}

The transformation ansatz asserts that the leading secular corrections to
the fermion mode function come from taking the expectation value, in
the graviton vacuum, of the transformed tree order solution,
\begin{equation}
\Psi^{\rm TA}_{i}(x) \equiv \Bigl\langle \Omega \Bigl\vert \Lambda_{ij}
u^{(0)}_{j}\Bigl(\omega^{0}_{~\rho} x^{\rho},\vec{k},s\Bigr) e^{i k_{j}
\omega^{j}_{~\sigma} x^{\sigma}} \Bigr\vert \Omega \Bigr\rangle \; .
\label{TAfermion}
\end{equation}
Here $\omega^{\mu}_{~\nu}$ is the general coordinate transformation 
which was worked out in section 2.2, and $\Lambda_{ij}$ is the associated
local Lorentz transformation constructed in section 2.3. In both cases 
the graviton fields are evaluated at the same point $x^{\mu} = 
(\eta,\vec{x})$ as the mode function. As with the photon case, this is
invalid because the transformations were constructed under the assumption
of constant graviton fields. Indeed, there is no transformation which 
would restore de Sitter for spacetime dependent graviton fields. However,
we evaluate (\ref{TAfermion}), and compare it with the exact result
(\ref{fermionresult}), to demonstrate that secular graviton corrections
are neither the same as infrared divergences, nor are they pure gauge.

Because only spatial graviton fields engender secular dependence 
(\ref{VEV}), we can neglect any graviton fields with temporal components
from $\omega^{\mu}_{~\nu}$ and $\Lambda_{ij}$. The result of doing this
for $\omega^{\mu}_{~\nu}$ was given in expression (\ref{omeganotime}).
From (\ref{LAMBDA}), and the expansion (\ref{Bexp}), we see that dropping
the temporal graviton fields makes a dramatic simplification in the
Lorentz transformation,
\begin{equation}
\Lambda_{ij} \longrightarrow \delta_{ij} \; .
\end{equation}
It follows that we need only the spatial plane wave factor already expanded
in (\ref{planewave}). Taking the expectation value using (\ref{VEV}) gives,
\begin{eqnarray}
\lefteqn{\Psi^{\rm TA}_{i}(x) = u^{(0)}_{i}(\eta,\vec{k},s) e^{i \vec{k} 
\cdot \vec{x}} + } \nonumber \\
&& \hspace{1cm} \frac{\kappa^2 H^2 \ln(a)}{16 \pi^2} \Bigl[3 i \vec{k} 
\!\cdot\! \vec{x} \!-\! k^2 x^2 \!+\! (\vec{k} \!\cdot\! \vec{x})^2\Bigr]
u^{(0)}_{i}(\eta,\vec{k},s) e^{i \vec{k} \cdot \vec{x}} + O(\kappa^4)
\; . \qquad \label{TAfermion}
\end{eqnarray}
This possesses the same exotic spatial dependence as the transformation
ansatz prediction for the photon (\ref{TAphoton}), and it is equally 
discordant with the exact computation (\ref{fermionresult}).

\section{Discussion}

The cosmological perturbations generated by primordial inflation arise
because the mode functions of massless, minimally coupled scalars and
gravitons freeze in to nonzero constants (\ref{IRu}-\ref{IRv}) after 
their physical wave lengths have red-shifted beyond the Hubble radius. 
The fact that more and more modes reach this saturated limit as inflation 
proceeds is responsible for the appearance of secular loop corrections. 
This inevitable consequence of freezing-in has occasioned much angst and 
scepticism within the very same community which hails the tree order 
effect --- the generation of primordial perturbations --- as a triumph of 
inflation theory. This curiously contradictory attitude is often justified 
by arguing that because the spacetime dependence of very long wavelength 
metric perturbations cannot be discerned by a local observer, they should
be subsumed into a coordinate redefinition. The argument runs that secular
loop corrections are merely the effect of evaluating tree order, noninvariant 
correlators at these transformed coordinates. We refer to this belief as the
{\it transformation ansatz}.

There are two reasonable approaches to demonstrating the reality of 
secular loop corrections:
\begin{enumerate}
\item{Show that they appear even in the expectation values of invariant 
operators \cite{Tsamis:1989yu,Tsamis:2013cka}; and}
\item{Evaluate a tree order, noninvariant correlator at the appropriately
transformed coordinates and compare the result with secular loop corrections
computed in an exact computation.}
\end{enumerate}
We have here followed the second approach, taking as our points of comparison
two exact one loop computations of the quantum gravitational corrections to
the photon \cite{Leonard:2013xsa,Wang:2014tza} and fermion \cite{Miao:2005am,
Miao:2006gj} wave functions on de Sitter background. In neither case did
the leading secular dependence implied by the transformation ansatz ---
expressions (\ref{TAphoton}) and (\ref{TAfermion}) --- agree with the exact 
computation --- expressions (\ref{photonresult}) and (\ref{fermionresult}). 
The unavoidable conclusion would seem to be that the transformation ansatz is 
incorrect. That is not surprising because the ansatz was based on ignoring the 
spacetime dependence of the graviton field whereas the phenomenon of secular
growth derives precisely from the continual passage of modes from the 
spacetime dependent, sub-horizon form to the spacetime constant, super-horizon
form.

It is conceivable that some other ansatz can be devised to explain away 
secular dependence as the sceptics wish. There have been many, many expressions 
of the belief that it is pure gauge, and we confess to some frustration in 
extracting explicit and testable assertions from them. Much of the relevant 
literature seems based on confusing infrared divergences (which really should 
be pure gauge) with secular growth (which seems to be a physical effect). It 
is also characterized by imprecision about approximations (small is not the 
same as zero, $\epsilon = 0$ is not the same as $\epsilon \neq 0$,  
$\dot{\epsilon} = 0$ is not the same as $\dot{\epsilon} \neq 0$), and 
by poorly articulated principles which occasionally shade into mysticism. 
For example, why are time dependent but spatially constant quantities
unobservable? How can any local field become constant in an interacting,
four dimensional quantum field theory? Why do alleged proofs of these
assertions begin by specializing to unphysical models of inflation which
exclude the normal fields of the Standard Model and are guaranteed not 
to experience sufficient reheating? And why would it be a tragedy for 
the power spectra to be time dependent in their 8th significant figure? 

The degree of scepticism towards secular effects from quantum gravity is
sometimes difficult to fathom. For example, no one doubts that a homogeneous 
ensemble of gravitational radiation on flat space background would scatter 
light by an amount which increases the further (and hence the longer in 
time) the light propagates. This is the basis for pulsar timing measurements
of gravitational radiation. How can there be any doubt that inflationary 
gravitons have the same effect?

In formulating the transformation ansatz we have done our best to extract 
a clear and explicit enunciation of the gauge artifact belief which can be 
checked. However, it is possible that those sceptical about secular loop 
effects have some other analytic realization in mind. If so, we apologize 
for having misinterpreted their work, we invite them to enlighten us as to
its true meaning, and we propose that they check it against the explicit 
one loop results (\ref{photonresult}) and (\ref{fermionresult}).

\centerline{\bf Acknowledgements}

We are grateful for conversations and correspondence with S. Giddings,
G. Pimentel and M. Sloth. This work was partially supported by NSF grant 
PHY-1506513 and by the Institute for Fundamental Theory at the UF.

\end{document}